\documentclass[usenatbib]{mn2e} \usepackage{epsfig}

\newcommand{\etal}{et al.\ } 
\def\gtsima{$\; \buildrel > \over \sim \;$}
\def\ltsima{$\; \buildrel < \over \sim \;$} 
\def\be{\begin{equation}}
\def\ee{\end{equation}} 
\def\ba{\begin{eqnarray}}  
\def\ea{\end{eqnarray}}
\def\eg{{\frenchspacing\it e.g.}}  
\def\gtsima{$\; \buildrel > \over  \sim \;$}
\def\ltsima{$\; \buildrel < \over \sim \;$} \def\prosima{$\; \buildrel  
\propto
   \over \sim \;$} \def\gsim{\lower.5ex\hbox{\gtsima}}
\def\lsim{\lower.5ex\hbox{\ltsima}} \def\simgt{\lower.5ex\hbox{\gtsima}}
\def\simlt{\lower.5ex\hbox{\ltsima}}  
\def\simpr{\lower.5ex\hbox{\prosima}}
 
\begin{document}
\title[The effect of minihaloes on cosmic reionization] {The effect of
   minihaloes on cosmic reionization}

\author[B. Ciardi et al.]  {B. Ciardi$^{1}$, E. Scannapieco$^{2}$, F.
   Stoehr$^{3}$, A. Ferrara$^{4}$, I. T. Iliev$^{5}$, \& P. R.  
Shapiro$^{6}$\\
   $^1$ Max-Planck-Institut f\"{u}r Astrophysik, 85741 Garching,  
Germany\\ $^2$
   Kavli Institute for Theoretical Physics, UC Santa Barbara, Santa  
Barbara, CA
   93106\\ $^3$ Institut d'Astrophysique de Paris, 75014 Paris, France\\  
$^4$
   International School for Advanced Studies, Trieste, Italy\\ $^5$  
Canadian
   Institute for Theoretical Astrophysics, University of Toronto,  
Toronto, ON
   M5S 3H8, Canada\\ $^6$ Department of Astronomy, University of Texas,  
Austin,
   TX 78712-1083}

\maketitle \vspace {7cm}

\begin{abstract}

   One of the most debated issues in the theoretical modeling of
   cosmic reionization is the impact of small-mass
   gravitationally-bound   structures.  We carry out the first
   numerical investigation of the role of such   sterile `minihaloes',
   which serve as self-shielding screens of ionizing   photons.
   Minihaloes are too small to be properly resolved in current
   large-scale cosmological simulations, and thus we estimate their
   effects using a sub-grid model, considering two cases that bracket
   their effect   within this framework.  In the `extreme suppression'
   case in which minihalo   formation ceases once a region is
   partially ionized, their effect on cosmic reionization is modest,
   reducing the volume-averaged ionization   fraction by an overall
   factor of less than 15\%.  In the other extreme, in which minihalo
   formation is never suppressed, they delay complete   reionization
   as much as $\Delta z\sim 2,$ in rough agreement with the results
   from   a previous semi-analytical study by the authors.  Thus,
   depending on the details of the minihalo formation process, their
   effect on the overall progress of reionization can range from
   modest to significant, but the minihalo photon consumption is by
   itself insufficient to force an   extended reionization epoch.

\end{abstract}


\section{Introduction}

The study of reionization has recently witnessed an explosion of
observational results. Since the advent of the new century, more than
10 quasars with $z>5.7$ have been discovered \citep{White03,Fan04},
indicating a strong increment of the neutral Lyman-$\alpha$ optical
depth \citep[GP]{GP65}   with increasing redshift, often interpreted
as the end of reionization.  At   the same time observations of Cosmic
Microwave Background (CMB)   anisotropies have constrained the total
optical depth of CMB photons to Thomson   scattering in the range
$\tau_e=0.17 \pm 0.04$, where the quoted uncertainty depends   on the
analysis technique employed \citep[]{Kogut03,Spergel03}.  Independent
of the details, this optical depth can be converted into a
reionization   redshift that is $\simgt 10$. These tests,together with
other probes of the high-$z$ universe as, such as Ly$\alpha$ emission
and Gamma Ray Bursts  (see Ciardi \& Ferrara 2005 and references
therein), provide an invaluable   set of observational
information. But only measurements of the 21~cm line emission from the
neutral intergalactic medium (IGM) and collapsed haloes  will be able
to map the temporal and spatial evolution of the process itself
\citep[\eg][]{MMR97,Shaver99,Tozzi00,ISFM02,ISMS03,CiardiMadau03,FSH04}.

Numerical simulations and semi-analytical studies have struggled to
match this early reionization \citep[for a review
see][]{BarkanaLoeb01,CiardiFerrara05}, which requires an enhanced
ionizing photon emission at high redshift,   while delaying complete
overlap until $z\sim6$. Some models have even included scenarios such
as double reionization through very massive, metal-free   stars at
high redshift, and more normal stars at lower redshift
\citep[\eg][]{Cen03,WyitheLoeb03}, or early partial reionization due
to   a decaying particle or X-ray photons, followed by later
full-reionization   by normal stars
\citep[\eg][]{Oh01,CK04,HH04,RicottiOstriker04}.  More conservatively,
\citet[][CFW hereafter] {CFW03} have shown that normal   stars with a
slightly top-heavy Initial Mass Function (IMF) could suffice to
produce an early reionization epoch, in agreement with the
observations of the   {\em Wilkinson Microwave Anisotropy Probe
(WMAP)} satellite \citep[\eg][]{Kogut03}.

The basic conclusion one can draw from these models is that boosting
the ionizing photon production to the level required for early
reionization   is not a problem. Nevertheless, some tension might
arise when the models are   combined with the results from the
Gunn-Peterson effect, as the reionization   history must be tuned to
(i) produce a large optical depth (to comply with the   {\em WMAP}
constraint), and (ii) to complete reionization at a relatively   low
redshift $\sim 6$ (as suggested by the GP data).  Note, however, that
this tension might be simply an artifact of the limited statistics of
QSO absorption lines at these high redshifts
\citep[\eg][]{ChoudhuryFerrara05}.  A scenario satisfying the above
requirements could be one in which cosmic reionization starts at very
high redshift and then goes through a long phase of partial
ionization lasting approximately up to $z \sim 6$.

Such a long-lasting phase might be caused by the presence of
small-scale, gravitationally-bound structures \citep[][hereafter
ISS]{ISS05}.  In hierarchical theories like the cold dark matter (CDM)
model, the   smallest structures are the first to collapse and
virialize, and in order to   form stars they must radiate their virial
energy.  However, in a purely atomic gas   of primordial composition,
radiative cooling is ineffective below $10^4$~K.  At these
temperatures molecular hydrogen is the main coolant. But   H$_2$ is
easily dissociated by UV photons in the Lyman-Werner bands between
11.2 and 13.6~eV, which are copiously  produced by the first stars
\citep[\eg][]{HRL97,CFGJ00}.  On the other hand, an early X-ray
background could provide enough free electrons to promote H$_2$
formation. Although the relative strength of   the above feedback
effects is controversial, the positive feedback from   X-rays would
not be able to balance UV photodissociation in the smallest objects
\citep[\eg][] {HAR00,GB03}. Thus, we can expect a population of
sterile   ``minihaloes'' (MHs), which serve as self-shielding screens
of ionizing photons and increase the local   effective recombination
rate of the gas. This prolongs the partial reionization   phase and
delays the end of the reionization process.

When an intergalactic I-front encounters a minihalo during cosmic
reionization, the minihalo traps the I-front, converting it from a
supersonic, weak R-type front to a subsonic, D-type front, which
photoevaporates the minihalo gas.  The first detailed studies of this
interaction, using   numerical gas dynamics simulations with radiative
transfer, were reported in \citet[][hereafter, SIR]{SIR04} and
\citet[][hereafter, ISR]{ISR05}.    Since the gas in MHs is much
denser than in the surrounding IGM, the   recombination rate during
this process is similarly increased.  Furthermore, MHs were   so
abundant during reionization that a sight-line between any two given
ionizing sources would have passed through many of them.  This means
that MHs   have the potential to trap intergalactic I-fronts before
they overlapped, and   suggests that they may have drastically
increased the global consumption of   ionizing photons during
reionization \citep{HAM00,BL02,S01,SIRM03,SIR04,ISR05}.

While approximate schemes to account for the effect of MHs on photon
consumption during reionization had been previously proposed
\citep[\eg][]{HAM00,BL02}, they did not have the advantage of
incorporating the results of the detailed numerical simulations of
ionization front-MH interactions reported in SIR and ISR. These last
studies not only   highlighted the importance of accounting for the
gradual ``peeling away" of neutral   MH gas by ionization fronts, but
also provided detailed fits to the total   number of ionizing photons
absorbed as a function of MH mass, source flux level, and redshift.
These fits then serve as a sound foundation for more detailed
analytical treatments \citep[such as][]{ISS05} and numerical
investigations, that is the subject taken up here.

In the semi-analytical treatment in ISS, 
the rate of expansion of the
I-fronts around individual source halos is modified to include
the effect of MHs.  This was done by generalizing
the I-front continuity jump condition from Shapiro
and Giroux (1987) to include the consumption of extra ionizing
photons by MHs, as computed in 
SIR and ISR.  The time evolution of the total ionized
volume fraction of the universe was then calculated by summing
these time-varying ionized volumes created by each source, over the
statistical distribution of source halos in the universe.
 It was found that MHs increased the number of ionizing
photons required to complete reionization by as much as a factor
of 2, delaying completion of reionization by a redshift interval
$\Delta z \sim 2.$  
The fact that the mass fraction collapsed into MHs grows over time, 
suggested that the extra photon consumption by MHs would be more
efficient at  slowing the advance of I-fronts at late times.  
Instead, we found that the spatial clustering around each
ionizing source kept the density of MHs fairly constant,
where they were actually encountered by the I-fronts.
As a result, MHs decreased the availability
of photons to ionize the IGM at all epochs, delaying
reionization, without extending its duration sufficiently
to reconcile the GP and {\em WMAP} constraints.  
However, a more significant effect in that direction was found when the 
rising value of the small-scale clumping factor of the IGM was also taken
into account.

Here we focus on a complementary approach to this problem,
in which the extra photon consumption by MHs is incorporated as a
``subgrid'' correction in a numerical simulation of reionization.
We use N-body results to identify the ionizing sources
and the evolving IGM gas density field, a semi-analytical
approach to identify the local density of MHs,
and numerical radiative transfer techniques to propagate 
the I-fronts in  three dimensions.
The  advantage of the numerical treatment is that it
is liberated from the assumptions of spherically-averaged I-front
propagation and a reliance on analytical models for clustering
and bias.   The numerical treatment, for example, is better equipped to
account for the effects of clustered sources, which can result in
H II regions powered by multiple sources, either simultaneously or
sequentially.   On the other hand, the
semi-analytical approach has the advantage that it is not limited by
finite numerical resolution and dynamic range, while these issues are
inherent in the numerical treatment.  
Our hope is that, by offering these two distinct approaches, we can
gain confidence in their answers, to the extent that they agree,
remain cautious, to the extent that they do not, and gain more insight into
the important effects than either one affords, on its own.

The main goal of this paper is to assess if the photon consumption
provided by MHs can result in a slow and gradual reionization, thus
offering a   natural explanation for the high value of $\tau_e$ and,
at the same time,   account for the late reionization epoch hinted by
the GP opacity.  The paper is   organized as it follows. In
\S~\ref{reion} and~\ref{mh} we describe the   simulations of cosmic
reionization in the absence of MHs and the physics of MHs
photoevaporation, respectively. In \S~\ref{rt} we discuss the
implementation of photoevaporation physics in the simulation of
reionization. In \S~\ref{results} we present our results
and in \S~\ref{conclusions} we summarize our conclusions.

\section{Simulations of Cosmic Reionization}
\label{reion}

In this study we recompute the simulations of cosmic reionization  
described in
\citet[][hereafter CSW]{CSW03} and CFW, now including the effect of  
unresolved
MHs. In this Section we briefly summarize the main features of the
simulations. We refer the reader to the above papers for more details.

The simulations, performed with the N-body code {\tt GADGET}  
\citep{SYW01},
follow the evolution of an ``average'' region of the 
Universe\footnote{Throughout this study we assume a $\Lambda$CDM  
cosmology   with
($\Omega_m,\Omega_\Lambda,\Omega_b,h,\sigma_8,n)=(0.3,0.7,0.04,0.7,0.9,1
)$, where $\Omega_m$, $\Omega_\Lambda$, and $\Omega_b$ are the total
matter, vacuum, and baryonic densities in units of the critical
density, $h$   is the Hubble constant in units of 100
$km\,s^{-1}Mpc^{-1}$, $\sigma_8$ is   the standard deviation of linear
density fluctuations on the $8   h^{-1}{\rm Mpc}$ scale at present,
and $n$ is the index of the primordial power   spectrum
\citep[\eg][]{Spergel03} with the transfer function taken from
\citet{EBW92}.}.  The ``re-simulation'' technique
\citep[\eg][]{TBW97} has been used to follow at higher resolution the
dark matter distribution   within an approximately spherical region of
diameter $\sim 50 h^{-1}$~Mpc   included in a much larger volume
\citep[$479h^{-1}$~Mpc on a side;][]{YSD01}.  The position and mass of
dark matter haloes were determined with a friends-of-friends
algorithm, while gravitationally-bound substructures   within the
haloes were identified with {\tt SUBFIND} \citep{Springel01} and
were used to build the merging tree for haloes and subhaloes. The
smallest   resolved haloes (which start to form at $z\sim 19$) have
masses of $M \sim {\rm few}\times10^9$~M$_\odot$, and the galaxy
population was modeled via the semi-analytic technique of
\citet{Kauffmann99}.  A catalogue of galaxies for each of the
simulation outputs was obtained, containing for each galaxy, among
other quantities, its position, mass   and star-formation rate
\citep[see][]{Stoehr03}.

Within the high-resolution spherical sub-region, a cube of comoving
side $L=20 h^{-1}$~Mpc was extracted to study the details of the
reionization   process, using the Monte Carlo radiative transfer code
{\tt CRASH} \citep[][CFMR hereafter]{MFC03,Ciardi01} to model the
propagation of ionizing photons   into the IGM.

The only inputs necessary for the calculation are the gas density
field   and ionization state, as well as the source position and
emission   properties, which are provided at each output of the galaxy
formation simulation   described above. In particular, the dark matter
density distribution is tabulated   on a mesh using a Triangular
Shaped Cloud (TSC) interpolation \citep{HE81}.  Assuming that the gas
distribution follows that of the dark matter, the   gas density in
each cell grid is set requiring that $\Omega_b=0.04$. A   number
$N_c=128^3$ of cells have been used.  To reduce the computational cost
of the radiative transfer calculation, all the sources inside a grid
cell are   grouped into a single source placed at the cell
center. Mass and luminosity conservation are assured.

\section{Minihalo Photoevaporation}
\label{mh}

In the simulations described in the previous Section, the smallest  
resolution
element is a grid cell, $156\,h^{-1}$ kpc on a side.  Minihaloes are  
thus well
below the resolution of the N-body simulation, and there are up to  
thousands
of them inside each cell. The aim of this paper then is to model the  
presence
and photoevaporation of MHs as sub-grid physics.

According to ISR, the total number of ionizing photons absorbed per  
minihalo
atom during photoevaporation can be expressed as a function of the  
minihalo
mass, $M$, redshift, $z$, and level of external ionizing photon flux,  
$F$, as
follows \be \bar\xi_{\rm MH}(M,z,F_0)=1+\phi_1(M_7)\phi_2(z)\phi_3(F_0),
\label{xi_fit}
\ee where $M_7=M/(10^7M_\odot)$, and \be F_0\equiv  
\frac{F}{\{10^{56}\,\rm
   s^{-1}/[4\pi (1\,{\rm Mpc})^2]\}}.  \ee For sources with a $5\times  
10^4$~K
blackbody spectrum, typical for O-stars, $\phi_1(M) \equiv 4.4 \,
(M_7^{0.334+0.023\lg M_7})$, $\phi_2(z)\equiv \left({1+z}\right)/10 $  
and
$\phi_3(F_0) \equiv F_0^{0.199 -0.042 {\rm lg} F_0}$.  By combining
equation~(\ref{xi_fit}) with the average number of minihaloes in a given
volume, we can compute statistically the average ionizing photon  
consumption
by minihaloes per total atom in this volume as 
\be \bar{\xi}\equiv
\frac{1}{\bar{\rho}_m}\int_{M_{\rm min}}^{M_{\rm max}} dM \,
\frac{dn(M,z,\delta)}{dM} \, M \, \xi(M,z,F_0),
\label{meanxi}
\ee where $\bar{\rho}_m$ is the mean cosmological mass density and
$dn(M,z,\delta)/dM$ is the number density of minihaloes in a region  
with an
overdensity $\delta = \rho_m/{\bar \rho_m}-1$.  Here the upper mass  
limit is
set by the requirement that MHs are not able to cool atomically, that is
$T_{\rm vir, max} = 10^4$K or $M_{\rm max} = 2.8 \times 10^9  
(1+z)^{-3/2}
M_\odot$ in our cosmology.  Similarly, the lower mass limit, $M_{\rm  
min}$, is
taken to be the (instantaneous) 
Jeans mass in the cold neutral medium. 
Note that we neglect
the finite time delay  for the gas to respond  hydrodynamically
to cooling \citep{GH96}.
Accounting for this effect 
would raise the effective ``filtering scale'' in  
neutral regions, decreasing  the impact of minihalos somewhat.

Adopting an extended PS approach \citep[]{LC93}, we approximate the  
biased
number density of minihaloes, $dn(M,z,\delta)/dM$, as \be \frac{dn}{dM}  
=
\frac{\bar{\rho}_m}{M} \left|\frac{d \sigma^2(M)}{d M} \right| f  
\left[1.68
   D^{-1}(z)-\delta_L(\delta),\sigma^2(M) \right],
\label{dndMdelta}
\ee where $D(z)$ is the linear growth factor at a redshift $z$,  
$\sigma^2(M)$
is the variance of linear fluctuations within a sphere containing a  
mass $M$,
and 
\be 
f(\delta_L,\sigma^2) \equiv \frac{\delta_L }{\sqrt{2 \pi}  
\sigma^{3}}
\exp\left(-\frac{\delta_L^2}{2 \sigma^2} \right).  
\ee 
Finally,
$\delta_L(\delta)$ is the linear overdensity corresponding to a  
nonlinear
overdensity of $\delta$.  If $\delta \geq 0 $ these quantities can be  
related
by the standard top-hat collapse model in terms of a ``collapse  
parameter"
$\theta$ as: \be \delta = \frac{9}{2} \frac{(\theta - {\rm sin} \,  
\theta )^2}
{(1 - {\rm cos} \, \theta)^3} -1, \ee and \be \delta_L =
\frac{3}{5}\left(\frac{3}{4}\right)^{2/3} (\theta - {\rm sin} \,
\theta)^{2/3}.  \ee In underdense regions, $\delta$ and $\delta_L$ can  
be
related by the solution given by \citet{H77} \citep[see also][]{FP01}:  
\be
\delta =\eta(\delta_L)^{-3} - 1, \ee where \be \eta(\delta_L) =
-\frac{\delta_L}{3}+ \exp[-0.01 \, (4.73 \delta_L^2+0.83 \delta_L^3+0.10
\delta_L^4)], 
\label{etaL}
\ee is the ratio of the comoving size of a perturbation  
to its
initial comoving size.  
Note that eq.\ (\ref{dndMdelta}) is a Lagrangian number density,
which is normalized by the mean density $\bar \rho_m$ in eq.\ (\ref{meanxi})
rather than an Eulerian density, which would have 
been normalized by $(\delta+1) \bar \rho_m$ instead.

Using eqs.\ (\ref{dndMdelta})-(\ref{etaL}) we can then 
re-write equation~(\ref{meanxi}) as \be \bar{\xi}(z,\delta,F_0)= f_{\rm
   coll,MH}(z,\delta) + \phi_3(F_0) I(z,\delta),
\label{xi_ps}
\ee where \be f_{\rm coll,MH}(z,\delta)\equiv \frac{1}{\bar\rho_m}
\int_{M_{\rm min}}^{M_{\rm max}} dM \frac{dn(M,z,\delta)}{dM}M,
\label{fcoll}
\ee is the collapsed mass fraction in minihaloes and \be I(z,\delta)  
\equiv
\frac{1}{\bar\rho_m} \int_{M_{\rm min}}^{M_{\rm max}} dM
\frac{dn(M,z,\delta)}{dM}M\phi_1(M)\phi_2(z), \ee such that $\phi_3(F_0)
I(z,\delta)$ is the {\em extra} number of ionizing photons absorbed per
minihalo hydrogen atom.

\section{Reionization Simulations with Minihaloes}
\label{rt}

In this Section we discuss the implementation used to include the  
physics of
the MHs (\S~\ref{mh}) into the simulation of cosmic reionization
(\S~\ref{reion}). In general, the galaxy formation simulations and the
radiative transfer calculations are performed as described in \S~3 of  
CSW, and
only some of the quantities are modified to include the mean  
contribution of
MHs, as follows.

The gas density field provided by the galaxy formation simulations is  
split
into a diffuse component (the IGM) and MHs, according to the fraction  
of gas
collapsed into MHs, $f_{\rm coll,MH}$ (equation~\ref{fcoll}). That is,  
if
$n_{\rm H}$ is the hydrogen number density in a given cell, only a  
fraction
$n_{\rm H}(1-f_{\rm coll,MH})$ is assigned to the diffuse gas  
component.  From
$f_{\rm coll,MH}$ we can derive the total mass of MHs in the cell as:  
\be
M_{\rm MH}= \rho_m \Delta l^3 f_{\rm coll,MH}, \ee where $\rho_m$ is  
the gas
density in the cell and $\Delta l$ is its linear physical dimension.  
For our
reference simulation $f_{\rm coll,MH}(z,\delta)$ is calculated at
$(z_{cr},\delta_{cr})$, corresponding to the first photon package  
crossing of
the cell and never increased thereafter, i.e. once a cell has been  
crossed by
photons and fully or partially ionized, no new MHs are allowed to  
subsequently
form in the cell. In almost all cases, complete ionization of a given  
cell
occurs within $10^6$ years of the initial illumination.  We call this
reference case `extreme suppression' and it places a lower limit to the  
effect
of minihaloes on reionization.

On the other extreme, we consider the case in which we allow MHs
formation in partially ionized and recombined regions. In this case,
$f_{\rm   coll,MH}$ is updated at each time step and it never becomes
zero (we call this the   `no suppression' case). This is meant to
put an upper limit on the effect of MHs, but the formation of sterile
MHs in highly ionized regions is  not supported by any strong physical
motivation. In any case, although the impact of feedback on the
formation of primordial, small-mass objects has been investigated by
several authors \citep[\eg][]{Gnedin00,OhHaiman03,OShea05} a consensus
has not yet been reached.  In fact, an early X-ray background
would heat the  IGM before ionization from softer photons occurs and
may provide an entropy floor even stricter than our ``no suppression''
case.  As the presence and effectiveness of this mechanism remains
highly uncertain however \citep[see \eg][]{OhHaiman03,KM05},  we
can cautiously consider our calculations as providing reasonable
estimates of as upper and   lower limits on the impact of
MHs on cosmic reionization.

To estimate the fraction of ionizing photons absorbed by MHs, we  
evaluate the
optical depth of a given cell due to minihaloes alone, $\tau_{\rm MH}$,  
as:
\be \tau_{\rm MH}=0.56 \sigma(\nu) n_{\rm H} f_{\rm coll,MH} \Delta l.   
\ee
Here $\sigma(\nu)$ is the photoionization cross section at a frequency  
$\nu$
and the factor $f=0.56$ accounts for the average photon path length  
through a
cell (CFMR). The total optical depth of a cell includes the  
contribution of
both the diffuse gas, $\tau_{\rm diff}$, and MHs, $\tau=\tau_{\rm diff}+
\tau_{\rm MH}$, with: \be \tau_{\rm diff}=0.56 \sigma(\nu) n_{\rm  
HI}(1-f_{\rm
   coll,MH})\Delta l.  \ee In each cell along the path of the photon  
packet,
the value of the hydrogen ionization fraction is updated according to  
eq.~(10)
of CFMR. In this case, though, the number of photons deposited in the  
diffuse
component of the cell is replaced by $N_\gamma {\rm exp}(-\tau_{\rm  
MH})$, to
take into account the photons absorbed by the MHs. Here $N_\gamma$ is  
the
number of photons contained in the monochromatic packet that  
illuminates the
cell. Any photons in the package not absorbed by either the diffuse gas  
or the
MHs in the cell are passed onto the next cell.

To determine when the MHs in a given cell should be considered  
completely
photoevaporated in the `extreme suppression' case, we proceed as  
follows.  If
$(N_{\gamma, {\rm MH}})_i$ is the number of photons absorbed by MHs in  
a given
cell at time step $i$ of the simulation, the fraction of MHs mass that  
these
photons photoevaporate is: \be ({\cal F}_{\rm MH, evap})_i=  
\frac{(N_{\gamma,
     {\rm MH}})_i}{n_{\rm H}\Delta l^3 \bar{\xi}}, \ee where $\bar{\xi}$  
is
given by equation~(\ref{xi_ps}). The mass evaporated from the  
minihaloes,
$M_{\rm MH}{\cal F}_{\rm MH, evap}$, is transferred to the IGM and  
$f_{\rm
   coll,MH}$ is modified accordingly. Note that, as we assume that once  
the
cell has been illuminated no new MHs form, at each step only the flux
dependent coefficient in equation~(\ref{xi_ps}), $\phi_3(F_0)$, is  
updated
according to the value of the current flux reaching the cell, $F$,  
while the
rest is kept constant at $(z_{cr},\delta_{cr})$. $F$ is given by: \be
F=\frac{N_\gamma}{4\pi d^2 \Delta t },
\label{flux}
\ee where $\Delta t$ is the time elapsed since a photon packet has gone
through the cell and $d$ is the distance from the source. Once the sum  
$\sum_i
({\cal F}_{\rm MH, evap})_i$ becomes equal to unity, we set $f_{\rm
   coll,MH}=0$ and consider the minihaloes gone thereafter. 
Only the number of photons necessary to  
complete photoevaporation are absorbed 
during the final step in which this condition is met.

In the `no suppression' case instead, formation of minihaloes can
occur   at any time. For this reason, all the quantities in
equation~(\ref{xi_ps}) are   always updated. In particular, at each
time step, we first calculate how   much new mass has gone in MHs and
update $f_{\rm coll,MH}(z,\delta)$   accordingly. We then subtract the
evaporated mass from the MHs and transfer it to the IGM.  In this
case, when the sum $\sum_i ({\cal F}_{\rm MH, evap})_i$ becomes equal
to unity, new MHs are still allowed to form.

Finally, a slightly different approach is used for MHs inside a
computational cell containing an ionizing source (see Appendix).

\section{Results}
\label{results}

\subsection{Reionization histories}
\begin{figure}
   \psfig{figure=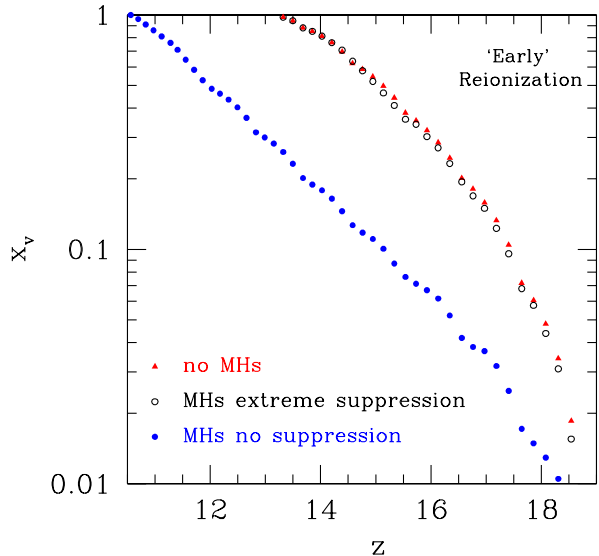,width=0.55\textwidth}
\caption{\label{ionfrac} Evolution of the volume-averaged ionization
   fraction for three different simulation runs (see text for details):  
run
   with no MHs (triangles), run with MHs and `extreme suppression' of  
their
   formation (open circles), and run with MHs and `no suppression' of  
their
   formation (filled circles). Results are for an `early' reionization  
case.}
\end{figure}

We quantify the impact of MHs on the progress of cosmic reionization
by computing the volume-averaged ionization fraction, $x_v$, as a
function   of redshift, as shown in Figure~\ref{ionfrac}. In our
reference run (open circles) we adopt an emission spectrum typical of
metal-free stars, a   mildly top-heavy Larson IMF with characteristic
mass of 5~M$_\odot$ and an   effective escape fraction of ionizing
photons $f_{esc}=20$\%.

 From a comparison with the analogous run in the absence of MHs
(triangles; L20, `early' reionization case in CFW), it is clear that
the presence   of MHs reduces the average ionization fraction most at
early times, where the difference is $\approx$ 15\%, while at later
times the difference is   $\lsim$ 10\%.  This reflects the fact that
less than 20\% of the emitted   photons in our model are absorbed by
MHs at high redshifts, and this fraction   decreases with decreasing
redshift. Of the above percentage only $\lsim$ 5\% is   absorbed by
MHs in the source cells. This is mainly because there are only $\sim
150$ (4000) source cells at $z\sim 19$ (10), while there are about $2
\times   10^6$ cells in the simulation. In both cases reionization is
complete by $z_{ion}\sim 13$.

\begin{figure}
   \psfig{figure=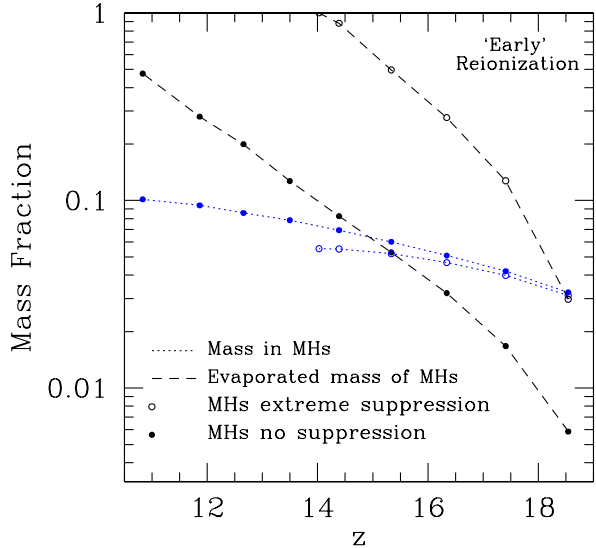,width=0.55\textwidth}
\caption{\label{mass} Redshift evolution of the fraction of total mass
   that is in MHs (excluding  photoevaporation, dotted lines) and
   the fraction of mass in MHs which is photoevaporated (dashed
   lines). Filled (open) circles indicate a case   of `no suppression'
   (`extreme suppression') of MHs formation.  Results are   for an
   `early' reionization case.}
\end{figure}

In Figure~\ref{mass} we show the evolution of the MH collapsed
fraction (ignoring  photoevaporation, dotted line, open circles) and
the   fraction of photoevaporated MH mass (dashed line, open circles).
The MHs collapsed fraction increases with decreasing redshift, until
it reaches a plateau, corresponding to the   epoch at which all the
cells in the simulation have been crossed by photons and   no more MH
formation is allowed. Note that the total mass in the simulation box
should be constant, but small fluctuations are possible as it is cut
from a   spherical region that is embedded in a much larger volume.
The fraction of photoevaporated mass of MHs increases with time, until
a value of the   order of unity is reached at $z \sim 14$.  Note that
reionization of the IGM and complete MHs photoevaporation do not
necessarily coincide, as there   could be cells in which the
ionization fraction of the diffuse gas is $\sim 1$,   but some MHs
still survive photoevaporation, or vice-versa.

The `no suppression' case (filled circles in Fig.~\ref{ionfrac}) shows
a much stronger effect due to MHs, which absorb 80-90\% of the
ionizing   photons.  As a result, reionization is delayed by $\Delta
z\sim 2$. These results are similar to those obtained in
ISS, which was largely focused on the impact of MH around individual sources, 
which were later summed together to give a rough estimate of the total
MH impact on reionization.   This means that each source in ISS saw a full 
compliment of MH, a picture very much like the `no suppression' case
studied here.

In   Figure~\ref{mass} the fraction of mass in MHs (dotted
line, filled circles) follows a  trend similar to the `extreme
suppression case', but now the abundance of MHs increases at late
times, and at early times the fractional abundance of   MHs is
slightly higher.  At all redshifts, the fraction of mass in MHs that is
photoevaporated (dashed line, filled circles) is much smaller 
and increases  more slowly than the `extreme suppression'
case. This is because MHs are  continuously re-formed, while in the
previous case no re-formation follows the almost instantaneous
photoevaporation.  As a consequence of re-formation, MHs survive also
after complete  reionization of the diffuse gas.  If this is the
case, their presence would still be visible in terms of,  \eg, a
contribution to the optical depth to ionizing photons and to the 21~cm
emission line.

Finally, we have also explored the `late' (S5) reionization case of  
CFW, in
which a Salpeter IMF and $f_{esc}=5$\% are adopted.  Carrying out an  
`extreme
suppression' simulation, we find that the effect of MHs is  
qualitatively and
quantitatively similar to our `early' reionization model in which no MH
reformation was allowed.  Less than 20\% of photons are absorbed by MHs  
at any  redshift.
This may reflect the fact that our MH collapsed fraction did not increase
much with decreasing redshift above the level obtained at the higher
redshifts of our early reionization case plotted in Figure~\ref{mass}.
Here we note that the MH collapsed fraction is somewhat
lower for this study than it was in ISS, due to our choice of 
the unknown small-scale power spectrum and, to a lesser extent,
our coarse-grain density field. We discuss their impact on
our results in Sec. 5.3.

\subsection{Thomson Scattering Optical Depth}

 From the above reionization histories it is possible to derive the  
expected
optical depth to electron scattering, $\tau_e$, as: \be  
\tau_e(z)=\int_0^z
\sigma_T n_e(z') c \left\vert \frac{dt} {dz'} \right\vert dz',
\label{tau}
\ee where $\sigma_T=6.65 \times 10^{-25}$~cm$^2$ is the Thomson cross  
section
and $n_e(z')$ is the mean electron number density at $z'$. Prior to  
complete
reionization, $n_e$ is obtained from the simulations at each redshift as
$n_e=\sum_i^{N_c} x_i n_i(1-f_{\rm coll,MH})/N_c$, where the sum is  
performed
over all the cells and $x_i$ is the ionization fraction in cell $i$.   
Once
reionization is completed, we simply assume complete hydrogen and He~I
ionization throughout the box. We also assume complete He~II  
reionization
after $z=3$. We find that the Thomson scattering optical depth is
$\tau_e=0.16$ for the `early' reionization case without MH reformation,
$\tau_e = 0.12$ for the 'early' case in which re-formation is allowed,  
and
$\tau_e = 0.10$ for the `late' case without re-formation.

\subsection{Model Uncertainties}
\label{approx}

In this Section we discuss the approximations and assumptions we  
adopted to
model the effect of MHs on the progress of reionization.

The MH photoevaporation simulations by ISR on which we based our model  
span a
range of fluxes $0.01<F_0<1000$.  Very close to the ionizing sources,  
however,
the normalized photon flux $F_0$ in equation~(\ref{flux}) can exceed  
1000. In
the framework of our galaxy model this occurs rarely, in less than 2\%  
of the
cases, and thus it affects only a small fraction of the volume of the
simulation. In such cases we set $F_0=1000$ in equation~(\ref{xi_fit}),  
which
is a reasonable approximation since at high fluxes the MHs photon  
consumption
dependence on the external flux becomes relatively weak. However, under
different assumptions about the sources lifetimes, e.g. short-lived  
sources,
the photon consumption by minihaloes would be higher.  In a higher  
percentage
of cases ($\sim 10$\%), $F_0$ becomes smaller than 0.01. When the flux  
becomes
so low, relatively few additional photons per atom are needed to  
evaporate the
MHs and thus we assume that just one photon per atom is absorbed if  
$F_0 \leq
0.01$. Both of these approximations are small, but conservative in  
terms of
estimating the photon consumption by MHs.

For each output of the N-body simulations, the dark matter density
distribution (and consequently the gas distribution, which is assumed to
follow that of the dark matter) is tabulated on a mesh using a TSC
interpolation \citep{HE81}. The TSC interpolation has the advantage of
producing smoothed density values at the positions of the grid cells, as
required for the radiative transfer computation. However it is not  
adaptive
and density variations on sub-grid scales are thus smoothed out,  
reducing the
contrast of the dark matter density distribution. In CSW this effect is
discussed in terms of a comparison between the TSC interpolation and a
32-particle SPH kernel, which should better capture the density  
contrasts.
They found that the overall agreement for low and intermediate density  
is
reasonable, but the smoothing of the TSC scheme is clearly visible at  
high
densities, especially at the lower redshifts (Fig.~9 of CSW). This can  
lead to
an underestimate of the value of $\delta$ at $z\simlt 12$ and dark  
matter
densities $\rho \simgt 10^{-(3-4)}$~M$_\odot/h ({\rm kpc}/h)^{-3}$,  
depending
on $z$.  To estimate the error introduced by the TSC scheme, we  
computed the
ratio of the average number of extra photons absorbed per unit hydrogen  
atom
at $z=10$ in both the TSC and SPH scheme. This is \be \frac{\sum_i^N
   \phi_3(F_{0,i}^{\rm TSC}) {{\bar \rho}_m}(10) I(10,\delta^{\rm  
TSC}_i)} {
   \sum_j^N \phi_3(F_{0,i}^{\rm SPH}) {{\bar \rho}_m}(10)  
I(10,\delta^{\rm
     SPH}_i)}.  \ee In general, $\phi_3(F_{0,i}^{\rm TSC})$ and
$\phi_3(F_{0,i}^{\rm SPH})$ need not be equal and will depend in detail  
on the
formation and propagation of the ionization fronts in the simulations.
However, as a rough estimate we simply factor out these terms, leaving  
a ratio
of 1.030, or an error of $\sim$ 3\% due to smoothing.

Due to our coarse-grained density field the total minihalo collapsed  
fraction
in our computational volume obtained using our model (Fig.~\ref{mass})  
is
slightly lower than the global collapsed fraction obtained directly from
analytical estimates (e.g. using PS or ST mass functions), particularly  
at
later times when density fluctuations become more nonlinear.  
Higher-resolution
N-body simulations would be required to better calibrate the mass
function-local overdensity relationship better, but this goes beyond the scope of  
this
paper. Note that these effects are most severe in cells containing  
ionizing
sources, in which internal density contrasts are higher.

The semi-analytical treatment by ISS found that
the importance of MHs as consumers of
ionizing photons was higher if each source emitted its total lifetime supply
of ionizing photons in a short burst at higher luminosity than if it
emitted the same number of photons continuously over time at lower luminosity.
In the Monte-Carlo reionization simulations reported here, each source is
assumed to release its supply of ionizing photons spread out over the
time between time-slices of the density field which were provided by the
galaxy formation simulations. This results in
longer source lifetimes and smaller luminosities than the
fiducial case considered by ISS, involving short-lived sources.
It is possible, therefore, that the MH correction effect would have been
higher if we had, instead, assumed shorter source lifetimes.

As there are currently no direct constraints on the power spectrum of
density fluctuations at minihalo scales, our estimates require
a large, uncertain extrapolation.
Although the transfer function we adopted is consistent with
the simulations, assuming a different transfer function, such as
the often-used one by \citet{EH99}, can change the minihalo numbers
significantly.  For this particular application, the
collapsed fraction in minihaloes for the \citet{EH99} power spectrum is
larger than our fiducial model by almost a factor of two, implying a
similar correction factor to the global effect of minihaloes.
In fact, we have conducted a comparison simulation for the `extreme
suppression' case in which we use the \citet{EH99} power spectrum for
the minihalo component, and we obtain an ionization fraction which  
is $\sim$ 90\% of our fiducial run.
Since currently there are no observational constraints to distinguish
one of these transfer functions from the other, we have chosen to
remain consistent with our large-scale N-body simulations, but
again, this could result in a conservative estimate of the
minihalo photon consumption during reionization.

A similar uncertainty arises from our choice of an analytic form
for the MH mass function. Over the relevant redshifts, shifting from the
Press-Schechter mass function to a \citet{ST02}
form would decrease
the total mass in  minihalos by roughly a factor of 1.5.  However,
there is little  evidence to motivate the use of this more complicated
expression at very high redshifts.  In fact, the few simulations of
high-z small-scale structure formation that exist seem to agree 
reasonably well with the Press-Schechter approach \citep[]{S01,JH01,Cen04}.

The resolution of the simulations is not able to capture the IGM
density distribution on very small scales. Usually this is taken into
account,   both in numerical simulations and semi-analytic approaches,
through the so   called gas clumping factor. As the main goal of this
study is to discuss the   effect of MHs on the reionization process
through a direct comparison with the simulations described in CSW and
CFW, we have not included this effect.   Its inclusion would delay the
reionization process of an amount depending   on the adopted value or
expression for the clumping factor (see e.g. ISS).  It   should be
noted, however, that according to the results of ISS, any increase
of the global photon consumption due to minihaloes is combined with
the   increased consumption due to recombinations in the clumpy IGM,
often amplifying   their impact. Any complete treatment of the effect
of small-scale structures reionization should therefore include both
effects.

\section{Conclusions}
\label{conclusions}

We have investigated the impact of minihaloes on the reionization
process by means of a combination of high-resolution N-body
simulations (to  describe the dark matter and diffuse gas evolution),
high-resolution hydrodynamical simulations (to examine
the   photoevaporation of minihaloes), 
a semi-analytic model of minihalo formation (to follow their biased
distribution),  and the Monte Carlo  radiative
code CRASH (to follow the propagation of ionizing photons).  We have
studied the process assuming different parameters   that regulate the
ionizing photon emission and different prescriptions for   the
evolution of minihaloes. The main results discussed in this paper can
be summarized as follows.

$\bullet$ If minihalo formation in a cell is completely suppressed once  
the
first photon packet arrives at the cell, i.e. in partially or fully  
ionized
cells (`extreme suppression' case), we find that their effect on cosmic
reionization is modest, with a volume averaged ionization fraction that  
is only $\lsim 15\%$ lower than the one when minihaloes are ignored. 
Only   less
than 20\% of the total emitted ionizing photons are absorbed by  
minihaloes at
any redshift. Complete reionization is not delayed significantly by the
presence of minihaloes.

$\bullet$ If minihalo formation is not suppressed (`no suppression'  
case),
then up to 80-90\% of the emitted ionizing photons are absorbed by MHs  
and
complete overlap can be delayed by as much as $\Delta z\sim 2$.

The `no suppression' case is in good agreement with the results of          
the semi-analytical approach described in ISS. However, we find that the
impact of MHs is smaller than the estimates by \citet[]{HAM00} and \citet[]{BL02},
as these authors have overestimated the number of photons required to photoevaporate
a MH. In fact, \citet[]{BL02} use static models of clouds in thermal and ionization 
equilibrium without accounting for gas dynamics. For this reason they fail to capture
some essential physics and overestimate the number of recombinations inside MHs.
\citet[]{HAM00} employ hydrodynamic simulations to study photoevaporation, but 
they do not include radiative transfer. \citet[]{SIR04} and \citet[]{ISR05} (on which our
calculations are based) found that the effect of ignoring radiative transfer and
its feedback on the gas dynamics is to significantly overestimate the number of
ionizing photons required to evaporate a MH.

As discussed in \S~\ref{approx}, there are still uncertainties in our  
model,
some of which are conservative in terms of estimating the photon  
consumption per minihalo.  
However, the `no suppression' case certainly provides a  liberal
overestimate of the MH number density, which is likely to more than  
balance these effects.  On the other hand, the `extreme suppression'
case provides a lower limit to the average MH number density, since it
assumes that no new MHs form in any cell that has ever been exposed to
ionizing photons, no matter how weakly or long ago, and thus it is likely to
bracket the MH correction effect from below.
It is with some confidence, then, that we can limit the  
true impact of minihaloes to lie between the extreme cases considered here,
although future investigations will be necessary to limit this range further.
Thus, our results indicate that the photon sink provided by these structures
is not sufficient by itself to force a long-lasting phase of partial IGM
ionization. This agrees with the conclusion of ISS for the cases 
that neglected the small-scale clumping of the IGM.  
Thus the resolution of the apparent conflict between the
high electron scattering optical depth in the WMAP data 
and the low reionization redshift inferred by QSO 
absorption line experiments should rely on additional
processes, such as the evolving small-scale IGM gas clumping factor, 
feedback effects or a transition in the star formation mode.

\noindent
\section*{Acknowledgments}

We would like to thank an anonimous referee for his/her helpful comments.
The paper has been partially supported by the Research and Training  
Network
`The Physics of the Intergalactic Medium' set up by the European  
Community
under contract HPRN-CT-2000-00126. This work was partially supported by 
NASA Astrophysical Theory Program
 grants NAG5-10825 and NNG04GI77G to PRS. ES was supported by the National  
Science
Foundation under grant PHY99-07949. This work greatly benefited by the  
many
collaborative discussions made possible by visits by BC, AF, ITI, and  
PRS to
the Kavli Institute for Theoretical Physics as part of the 2004  
Galaxy-IGM
Interactions Program.

\appendix
\section{Minihaloes in source cells}

To account for the presence of MHs in computational cells that contain
ionizing sources, we must take a slightly different approach from that
described in \S4.  The reason is that at high redshift the ionizing  
sources
were rare peaks of the density distribution, and thus minihaloes were  
strongly
clustered around them. To account better for this clustering, we  
calculate the
total number of photons absorbed by the gas (both diffuse and MHs) in  
that
cell, $N_s$, using a simplified version of the analytical model we  
developed
previously in ISS. In the current simulations we group all sources  
within each
cell into a single source with a mass equal to the sum of the individual
masses. Note however that such multiple sources only occur at the latest
redshifts simulated and thus this approximation has only a minor effect  
on our
results.

 From ISS, the rate of expansion of the comoving volume about an ionized
source of mass $M_s$ is given by \be \frac{dV_I}{dt} =  
\frac{\dot{N}_\gamma
   (n_H^0)^{-1} - \alpha_B \, (1+z)^3 \, n_H^0 (V_I-V_0)}{1-f_{\rm
     coll}(M_s,r)+ \bar\xi_{\rm src}(M_s,z,F_0)},
\label{meanscreen}
\ee where $n_H^0$ is the average comoving number density of hydrogen,
$\alpha_B = 2.6 \times 10^{-13}$ cm$^3$ s$^{-1}$ is the case B  
recombination
coefficient for hydrogen at $10^4$ K, $V_0$ is the comoving volume  
initially
carved out by the material making up the sources, and \be f_{\rm
   coll,MH,src}(M_s,z,r) = \frac{1}{{\bar \rho}_m} {\int_{M_{\rm  
min}}^{M_{\rm
       max}} dM \, \frac{dn(M,z,r | M_s)}{dM} \, M \, },
\label{fcollsrc}
\ee and \ba \bar{\xi}_{\rm src}(M_s,z,F_0,r) &=& \frac{1}{{\bar \rho}_m}
\int_{M_{\rm min}}^{M_{\rm max}} dM \times \\&& \qquad
\frac{dn}{dM}(M,z,r|M_s) \, M \, \xi(M,z,F_0) \nonumber
\label{meanxisrc},
\ea are the MH collapse fraction and average number of extra absorbed  
photons
by MHs at a Lagrangian distance $r$ from a source of mass $M_s$ at a  
redshift
$z$ shining with a flux $F_0$.  Both $f_{\rm coll}(M_s,z,r)$ and
$\bar{\xi}_{\rm src}(M_s,z,F_0,r)$ make use of the biased number  
density of
minihaloes, forming at a distance $r$ from the source halo.  To  
calculate this
quantity, we make use of the analytical formalism described in detail in
\citet[]{SB02} \citep[see also][]{Petal98}.  Here the biased number  
density is
given as \be \frac{dn}{dM}(M,z,r|M_s) = \frac{dn^2}{dM  
dM_s}(M,z,M_s,z,r)
\left[\frac{dn}{dM_s}(M_s,z) \right]^{-1}, \ee where $dn/dM_s$ is the  
usual PS
mass function and $\frac{dn^2}{dM dM_s}(M,z,M_s,z,r)$ is the bivariate  
mass
function that gives the product of the differential number densities at  
two
points separated by an initial comoving distance $r$, at any two masses  
and
redshift.  Note that this expression interpolates smoothly between all
standard analytical limits, including those of \citet[]{MW96} and
\citet[]{LC93}, and it has also been carefully validated against  
simulations
\citep[]{ST05}.

Finally, we compute $N_s$ by integrating eq.~(\ref{meanscreen}) up  
until the
time $t_{\rm cell}$ at which the total {\em mass} contained within the  
ionized
region is equal to the mass contained within the source cell.  The  
number of
photons absorbed is then simply $N_s = t \dot N_\gamma,$ which is  
imposed by
adopting a lower effective escape fraction, such that the total number  
of
ionizing photons leaving the source cell is decreased.

\end{document}